\documentclass[onecolumn,showpacs]{revtex4}

\usepackage{graphicx}
\usepackage{dcolumn}
\usepackage{bm}

\input epsf

\begin{document}

\title{\Large Anisotropic Brane Cosmology with Variable $G$ and $\Lambda$}

\author{\bf Soma Nath}
\author{\bf Subenoy Chakraborty}
\email{subenoyc@yahoo.co.in}
\author{\bf Ujjal Debnath}
\email{ujjaldebnath@yahoo.com}

\affiliation{Department of Mathematics, Jadavpur University,
Calcutta-32, India.}

\date{\today}

\begin{abstract}
In this work, the cosmological implications of brane world
scenario are investigated when the gravitational coupling $G$ and
the cosmological term $\Lambda$ are not constant but rather there
are time variation of them. From observational point of view,
these time variations are taken in the form $\frac{\dot{G}}{G}\sim
H$ and $\Lambda \sim H^{2}$. The behavior of scale factors and
different kinematical parameters are investigated for different
possible scenarios where the bulk cosmological constant
$\Lambda_{5}$ can be zero, positive or negative.
\end{abstract}

 \maketitle

\section{\normalsize\bf{Introduction}}
There are two fundamental physical parameters namely the
gravitational coupling $G$ and the cosmological term $\Lambda$, in
the Einstein's theory of general relativity. Usually these
parameters are assumed to be constants. But recent experimental
and (or) observational evidences suggest that it is not unlikely
to have small variations [1] of these fundamental parameters.
However, alternative ideas about the variability of these
parameters have been started long ago. The idea of variable $G$
was first introduced by Dirac [2] in cosmological consideration
and then subsequently by Sciama [3], Jordan [4] and others.
Later, Brans and Dicke [5] proposed an extension (or modification)
of Einstein's theory of gravity by introducing a scalar field
$\phi$. In this theory, the Newton's gravitational constant $G$
is a variable and is related to the
scalar field $\phi\sim G^{-1}$.\\

The cosmological term $\Lambda$ on the otherhand, was originally
suggested by Einstein himself to obtain static solution of his
field equations. But after his realization that the universe is
expanding, he discarded the cosmological term. Afterwards, the
cosmological term has been introduced and subsequently rejected
several times for various reasons. But recent measurements of the
CMB anisotropy and the observations from type Ia supernovae demand
a significant and positive cosmological constant [6, 7]. Also
observations of gravitational lensing indicate the presence of a
non-zero $\Lambda$. However, to resolve the cosmological constant
problem (i.e., its observed value is about 120 orders of
magnitude below the value for the vacuum energy density predicted
by quantum field theory) a phenomenological solution namely a
dynamical $\Lambda(t)$ has been suggested by several authors [8],
arguing that $\Lambda$ relaxes its present estimate due to the
expansion of the universe. Also due to these recent observational
predictions the vacuum energy density
$\rho_{\Lambda}~(=\Lambda/8\pi G)$ has played an important role
and the cosmological constant problem has been shifted to the
{\it coincidence problem} namely why $\rho$ and $\rho_{\Lambda}$
happen to be of the same order of magnitude precisely at this
very moment. To resolve this naturalness several quintessence
models [9] have been proposed where the cosmological constant
becomes a time dependent quantity. Recently Bonanno et al [10]
(see also Shapiro et al [11]) have formulated a general frame work
for cosmologies in which both Newton's constant and the
cosmological constant are time dependent and have shown that
$\rho$ and $\rho_{\Lambda}$ are approximately equal in the late
universe. They have assumed that there exists an infrared
attractive fixed point for the renormalization group flow of the
(dimensionless) Newton's constant and cosmological constant
respectively. Vishwakarma [12] has proposed a variable $\Lambda$
proportional to $H^{2}$ ($H=$ Hubble parameter) and has showed a
good agreement with recent supernovae observations. Also this
choice provides expected large age of the
universe. \\

In this paper, we analyse a 5D brane world model with a time
dependent brane tension $\lambda$ and constant 5D cosmological
constant $\Lambda_{5}$. We shall show that this will lead to a 4D
effective theory with time dependent $G$ and $\Lambda$. Although
the fundamental theory is not yet known, the time dependent of
the self-energy of the brane $\lambda$ can be the result of some
higher dimensional effect, such as a particular bulk-brane
interaction, or the response of the brane to the change in the
inter-brane distance in a 2-brane world model. Thus here we
attempt  to make a ``phenomenological" study of the implications
of such a time-dependence of $\lambda$ on observable
4-dimensional parameters, such as $G$ and $\Lambda$, and its
consistency with current observational constraints.

\section{\normalsize\bf{Basic equations in Brane Cosmology }}
The idea of brane world scenario was proposed by Randall and
Sandrum [13]. They have shown that in a five dimensional
space-time (called {\it bulk}) it is possible to confine the
matter fields in a four dimensional hypersurface (called {\it
3-brane}). The effective action in the bulk is [14]

\begin{equation}
A=\int d^{5}x
\sqrt{-g_{_{5}}}~\left(\frac{1}{2\kappa^{2}_{5}}R_{5}-\Lambda_{5}\right)+\int_{\chi=0}d^{4}x
\sqrt{-g} ~\left(\frac{1}{2\kappa_{5}^{2}}K^{\pm}+\lambda +{\cal
L}_{matter}\right)
\end{equation}

Here any quantity  in the bulk is marked with subscript `5' with
$\kappa_{5}^{2} ( = \frac{8 \pi G_{5}}{c^{4}})$ as the 5D
gravitational coupling constant. The co-ordinates on the bulk are
denoted by $x^{a},~a=0,1,2,3,4$ while those on the brane as
$x^{\mu},~\mu=0,1,2,3$~. So $\chi=x^{4}=0$ the four dimensional
hypersurface, is defined as the brane world. The parameter
$\lambda$ stands for the brane tension which is assumed to be
positive to recover the conventional gravity on the brane. The
other parameter $\Lambda_{5}$ is the negative vacuum energy and
is the only source of gravitational field on the bulk while
$K^{\pm}$ are the intrinsic curvature on either side of the brane
(characterized by $\chi > 0$ or $< 0$) . Now the effective
Einstein equations on the bulk are [15, 16]

\begin{equation}
G_{ab}^{5}=-\Lambda_{5} g^{5}_{ab}+\kappa^{2}_{5}T^{5}_{ab
(\text{brane})}
\end{equation}

where $T^{5}_{ab (\text{brane})}$ (with $T^{5}_{ab
(\text{brane})}.n^{a}=0$) is the total energy-momentum tensor
(vacuum+matter) on the brane i.e.,

\begin{equation}
T^{5} _{ab(\text{brane})}=\delta (\chi)[\lambda g_{ab}+T^{(m)}
_{ab}]=\delta
(\chi)\delta^{\mu}_{a}\delta^{\nu}_{b}\tau_{\mu\nu}~~.
\end{equation}

Here $T^{(m)}_{ab}$ is usual matter energy tensor and the Dirac
delta function indicates the fact that matter is confined in the
space-like hypersurface $\chi=0$ (the 3-brane) with induced
metric $g_{ab}$. \\

From the Israel's junction conditions
$$
K_{\mu\nu}|_{\chi>0}-K_{\mu\nu}|_{\chi<0}=\kappa^{2}_{5}\left(\tau_{\mu\nu}-\frac{1}{3}g_{\mu\nu}\tau\right)
$$

Due to $Z_{2}$-symmetry of the bulk,
\begin{equation}
K_{\mu\nu}|_{\chi>0}=-K_{\mu\nu}|_{\chi<0}=\frac{1}{2}\kappa^{2}_{5}\left(\tau_{\mu\nu}-\frac{1}{3}g_{\mu\nu}\tau\right)
\end{equation}

Hence we obtain
$$
\tau_{\mu\nu}=\frac{2}{\kappa^{2}_{5}}\left(K_{\mu\nu}-g_{\mu\nu}K\right)
$$

So from (4) we have
$$
K_{\mu\nu}|_{\chi>0}=-\frac{1}{2}\kappa^{2}_{5}\left[T_{\mu\nu}^{(m)}-\frac{1}{3}g_{\mu\nu}(T^{(m)}+\lambda)\right]
$$

Then the effective Einstein equations on the brane are ([15]; see
also [17]) (choosing $8\pi=1, c=1$)

\begin{equation}
G_{\mu \nu}=-\Lambda g_{\mu \nu}+\kappa^{2} T^{(m)}_{\mu
\nu}+\kappa^{4 } _{5} S_{\mu \nu}-E_{\mu \nu}
\end{equation}
\\
(Note that Greek indices refer to brane world while English small
letters to bulk quantities).\\

The above Einstein equation has two correction terms to the
energy-momentum tensor namely the local correction term $S_{\mu
\nu}$ with expression

\begin{equation}
S_{\mu \nu}=\frac{1}{12}T^{(m)}
T_{\mu\nu}^{(m)}-\frac{1}{4}T^{\alpha (m)} _{\mu}T_{\nu
\alpha}^{(m)}+\frac{1}{24} g_{\mu \nu}(3 T^{\alpha \beta (m)}
T_{\alpha \beta }^{(m)}-T^{2 (m)})
\end{equation}

while the other correction term is the non-local effects from the
free gravitational field in the bulk with expression

\begin{equation}
E_{ab}=C_{a b c d} n^{c} n^{d}
\end{equation}
Here $C_{a b c d}$ is the usual Weyl tensor in the bulk, $n^{a}$
the unit normal to the hypersurface $\chi=0$ and the four
dimensional cosmological constant $\Lambda$ has the expression
\begin{equation}
\Lambda =\frac{\kappa^{2}_{5}}{2} (\Lambda _{5}+ \frac{\kappa
^{2} _{5} \lambda ^{2} }{6})
\end{equation}
with
\begin{equation}
\kappa^{2} =\frac{\kappa ^{4} _{5}  \lambda}{6}
\end{equation}

as the four dimensional gravitational constant. Now using the
conservation of the total matter energy-momentum tensor on the
brane namely
\begin{equation}
\tau^{\nu}_{\mu;~\nu}=0
\end{equation}
into the modified Einstein equations on the brane we have a
constraint on $S _{\mu \nu}$ and $E_{\mu \nu}$ as
\begin{equation}
(E^{\nu} _{\mu}- \kappa^{4} _{5} S^{\nu} _{\mu})_{;~\nu}=0
\end{equation}
Maartens has shown that $E_{\mu \nu}$ can be decomposed as [16,
17]
\begin{equation}
E_{\mu \nu}= -\frac{6}{\kappa^{2} \lambda} [U (u _{\mu} u_{\nu}
+\frac{1}{3}h_{\mu \nu}) +2u_{(\mu} Q_{\nu)} +P_{\mu \nu}]
\end{equation}
where the prefactor $\frac{6}{\kappa^{2} \lambda} $ (i.e.,
$\frac{\kappa_{5}^{4}}{\kappa^{4}}$) is introduced for
dimensional reasons (note that $\lambda^{-1}\rightarrow 0$  gives
the general relativistic limit) and the energy flux $Q _{\mu}$ and
energy stresses $P _{\mu \nu}$ has the following properties :\\

(i) $Q _{\mu}$ is a spatial vector (i.e., $ Q _{\mu} u^{\mu}=0$)\\

(ii)$P_{\mu \nu}$ is a spatial (i.e., $P_{\mu \nu} u^{\nu}=0$),
symmetric $(P_{(\mu\nu)}=P_{\mu\nu})$ and
trace-free ($P ^{\mu} _{\mu}=0$) tensor.\\

The scalar term $U$ is called dark energy density as it has the
same form as the energy momentum tensor of a radiation perfect
fluid. Also $u^{\mu}$ is the 4-velocity on the brane with $h
_{\mu \nu}$, the projection tensor orthogonal to $u^{\mu}$ on the
brane. Further, using the constraint equation (11) between $E
_{\mu \nu}$ and $S _{\mu \nu}$, it is possible to have evolution
equations for $U$ and $Q _{\mu}$ but not for $P _{\mu \nu}$. So
the system of equations on the brane is not, in general
closed.\\

As we are considering Bianchi models which have a simple
transitive three dimensional group of isometries $G _{3}$ on the
space-like hypersurfaces, so the cosmological models are spatially
homogeneous with proper time as the only dynamical variables.
Thus the metric on the brane can be written as [18]
\begin{equation}
ds ^{2}=- dt^{2} +\alpha^{2}_{i} (t) dx^{i}dx ^{i}
\end{equation}

Further, the physical parameters which are of observational
interest in cosmology are the following:\\

Expansion scalar: ~~~~~~~~~~~$\theta=3H= \sum ^{3} _{i=1} H
_{i}=\sum_{i=1}^{3}
\frac{\dot{\alpha}_{i}}{\alpha_{i}}$\\

Shear scalar:
\begin{equation}
\sigma ^{2}=\frac{1}{2} \sigma _{\mu \nu}\sigma ^{\mu
\nu}=\frac{3}{2} A H ^{2}
\end{equation}\\

Deceleration parameter: ~~~$ q = -\frac{(\dot{H}+H^{2})}{H^{2}}$\\

where the mean anisotropy parameter $A$ has the expression
$$A=\frac{1}{3} \sum ^{3} _{i=1} \left(1- \frac{H_{i}}{H}\right)^{2}$$

and $H_{i}$ and $H$ are the directional Hubble parameters and
mean Hubble parameters respectively.\\

Lastly, for Bianchi I and V model we get $Q _{\mu}=0$ but $P _{\mu
\nu }$ remains unrestricted. As there is no way of fixing the
dynamics of this tensor, we shall study the particular case $P
_{\mu \nu}=0$. Then from the constraint equation (11) the dark
energy density $U$ is a function of the proper time alone i.e.,
$U=U(t)$ and the evolution equation for $U$ takes the form
\begin{equation}
\dot{U} = - ~4HU
\end{equation}
which on integration gives
\begin{equation}
U = U _{0}~ V  ^{- \frac{4}{3}}
\end{equation}
where $V=\prod^{3}_{i=1} \alpha _{i}$ is the volume scale factor
and $U_{0}$ is an integration constant.

\section{\normalsize\bf Bianchi I Model }
The line element of a Bianchi I space-times generalizes flat FRLW
metric to the anisotropic case as

\begin{equation}
ds ^{2} = - dt ^{2}+ a_{1}^{2}(t)dx ^{2}+a_{2}^{2}(t)dy
^{2}+a_{3}^{2}(t)dz ^{2}
\end{equation}

We have taken perfect fluid as the matter on the brane with
expression
\begin{equation}
T_{\mu \nu}^{(m)}=(\rho +p)u _{\mu}u_{\nu}+p~ g_{\mu \nu}
\end{equation}

where the energy density $\rho$ and thermodynamic pressure $p$
satisfy the isothermal equation of state namely
\begin{equation}
p =(\gamma -1)\rho ,~~~1\leq \gamma \leq 2 .
\end{equation}

The combinations of the non-vanishing components of the Einstein
field equations on the brane (i.e.(5)) are [18]
\begin{equation}
3 \dot{ H} + \sum ^{3} _{i=1} H_{i}^{2}=\Lambda -
\frac{(3\gamma-2)}{2}\kappa^{2} \rho - \frac{(3\gamma -1)}{12}
\kappa^{4}_{5} \rho ^{2} +  \frac{U_{0}}{V^{\frac{4}{3}}}
\end{equation}
and
\begin{equation}
\frac{1}{V}\frac{d}{dt}(V H _{i})= \Lambda - \frac{(\gamma
-2)}{2}\kappa^{2} \rho - \frac{(\gamma-1)}{12}\kappa^{4}_{5}
\rho^{2}+\frac{U_{0}}{V^{\frac{4} {3}}}
\end{equation}

It is to be noted that so far we have not considered any
variation of the 4D physical parameters namely $G$ and $\Lambda$.
The variation of $G$ and $\Lambda$ shows a time variation of
brane tension $\lambda$ (see equations (8) and (9)), keeping 5D
quantities as invariant. From observational point of view it is
generally assumed that the time variation of $G$ can be expressed
in terms of the Hubble parameter $H$ as [19, 20]
$$
\frac{\dot{G}}{G}= 3gH  ~~~\text{i.e.,}~~ G=G_{0} V^{g}
$$

where the dimensionless parameter $g$ has present observational
bound $ |g| \leq 0.1 $ .and $G_{0}$ is constant of integration.\\

As from equation (9) $G \propto \lambda$ so the differential
equation in $\lambda$ has the same form namely,
\begin{equation}
\frac{\dot{\lambda}}{\lambda}=3gH~ {\text i.e.,}~
\lambda=\lambda_{0}V^{g}
\end{equation}
where $\lambda_{0}$ is integration constant. Also the integration
constants ($G_{0}$ and $\lambda_{0}$) are connected by the
relation (using eq.(9))
$G_{0}=\frac{\kappa_{5}^{4}}{6}\lambda_{0}$.\\

As a consequence, the energy conservation equation (10) becomes
\begin{equation}
\dot{\rho} + 3 \gamma \rho H = - \lambda_{0}  g  V ^{g-1} \dot{V}
\end{equation}
which on integration gives
\begin{equation}
\rho = D V ^{-(1+\beta)}
\end{equation}

with  $\lambda_{0}=F_{0}D$ and $1+\beta =- g=\frac{\gamma}{F_{0}+1}$.\\

Hence from observation $ \beta $ is restricted to $-1 < \beta
\leq -0.966$. Also from equation (9) the brane tension is
proportional to $G$ and we write
\begin{equation}
\lambda = \frac{D (\gamma -1 -\beta)}{(1+\beta)} V^{- ( 1 +
\beta)}
\end{equation}

Now using relations (8), (16), (22) and (23) in the field equation
(21) we have
\begin{equation}
\frac{1}{V} \frac{d}{dt} (V H _{i}) = \frac{\kappa _{5} ^{2}
\Lambda _{5}}{2} -  \frac{\kappa_{5}^{4} D ^{2} \gamma ^{2} \beta
}{12 (1 + \beta)^{2}} V^{-2(1 + \beta)}
+\frac{U_{0}}{V^{\frac{4}{3}}}
\end{equation}

So adding equation (26) for  $i=1,2,3 $ we get
\begin{equation}
\frac{1}{V} \frac{d}{dt} (V H ) = \frac{1}{V} \frac{d}{dt} (V H
_{i})
\end{equation}
which on integration gives
\begin{equation}
H = H_{i} + \frac{K _{i}}{V}
\end{equation}

The integration constants $K_{i} , ~i=1,2,3 $ are restricted by
the relation $ \sum ^{3} _{i=1}K_{i} =0 $.\\

Combining equation (26) with (27) we have the differential
equation in V (after integrating once) as (assuming $(1 +
\beta)\neq 0 $)
\begin{equation}
\dot{V ^{2}} = a V ^{2} + b V^{- 2\beta} + c V ^{\frac{2}{3}} + d
_{1}
\end{equation}
or in integral form
\begin{equation}
 t - t _{0} = \int \frac{dV}{ \sqrt{a V ^{2} + b V^{- 2\beta} + c V ^{\frac{2}{3}} + d
_{1}}}
\end{equation}

where ~~ $a=\frac{3}{2} \kappa _{5} ^{2} \Lambda_{5},~
b=\frac{\kappa _{5} ^{4}D ^{2} \gamma ^{2}}{4 (1+\beta)^{2}},~
c=9 U_{0}$~ and $(d_{1},t_{0})$ are integration constants.\\

Also the scale factors can be obtained integrating once equation
(28) and using equation (29) as
$$a_{i}=a_{i0} V^{\frac{1}{3}}~ \text{exp}
\left[\frac{K_{i}}{3}
\int\frac{dz}{\sqrt{a+bz^{2(\beta+1)}}}\right] ,~~i=1,2,3$$ with
$a_{i 0}$'s as integration constants and $z=V^{-1}~$. Further, if
we assume the variation of the cosmological term (as mentioned
earlier) as
\begin{equation}
 \Lambda =\xi H^{2}
\end{equation}
(with $\xi$, a function of time) then using (25) and (31) in
equation (8) we get a differential equation in $V$ as
 $$ \dot{V}^{2}= \frac{9}{2 \xi} \kappa^{2}_{5}
\Lambda_{5}V^{2}+
 \frac{3}{4\xi}\kappa^{4}_{5}D^{2}\frac{(\gamma
 -1-\beta)^{2}}{(1+\beta)^{2}}V^{- 2 \beta}$$

Now comparing with equation (29) we have
\begin{equation}
 a\left(1-\frac{3}{\xi}\right)V^{2(1+\beta)}+ {\frac{b}{\gamma
 ^{2}}}\left[{{\gamma^{2}-\frac{3}{\xi}(\gamma-1-\beta)^{2}}}\right]+cV^{2(\frac{1}{3}+\beta)}+d_{1}V^{2\beta}=0
\end{equation}
which is true for all time and for all values of $V$. As powers of
$V$ in the 3rd and 4th term are in no way compariable to the
first two terms for any choice of $\beta$ so we can choose the
integration constant $d _{1}=0$ and also $c=0$. We note that the
constant `$c$' is related to the non-local energy correction term
(corresponds to an effective radiation) which is constrained to be
small enough at the time of nucleosynthesis and it should be
negligible today .\\

Moreover as a solution of (32) we have (in addition to
$c=0=d_{1}$) three possibilities namely
\begin{equation}
 (i)~~ \xi =3~~~ \text{and}~~~ \gamma^{2}=(\gamma-1-\beta)^{2}~~~ \text{i.e.}~~~\gamma=\frac{(1+\beta)}{2}
\end{equation}
\begin{equation}
 (ii)~~ a=0~~~ \text{and}
 ~~~\xi=\frac{3(\gamma-1-\beta)^{2}}{\gamma^{2}}~~~~~~~~~~~~~~~~~~~~~~~~
\end{equation}
and

\hspace{.97in}$(iii)$ $\xi$ is a function of $V$.\\\\

It is to be noted that the first choice contradicts the
observational bounds for $\beta$ [19] (i.e., $-1<\beta\le
-0.966$), so only the second and the third possibilities will be
discussed in the following for all possible values of
$\Lambda_{5}$
namely, ~(i) $\Lambda_{5}>0$, ~(ii) $\Lambda_{5}<0$ ~~and  ~(iii) $\Lambda_{5} =0$.\\\\

{\bf Case~I}:~~  $\Lambda_{5}~(\text{i.e.},~ a) > 0$\\

Here we have the solution of $V$ from equation (30) as (the third possibility i.e.,
$\xi$ is a function of $V$, is the only choice)\\

\begin{equation}
 V^{(1+ \beta)} =\sqrt{ \frac{b}{a}}~ \text{sinh} [(1+ \beta) \sqrt{a}~ (t-t _{0})]
\end{equation}
which  simplifies to (choosing ~ $t _{0}=0$, so the big bang
singularity occurs at $t=0$)

\begin{equation}
V^{1 + \beta} = \frac{D~\kappa_{5}\gamma}{(1+\beta)\sqrt{6\Lambda
_{5}}}~ \text{sinh} [(1 + \beta ) \sqrt{\frac{3\Lambda_{5}}{2}}~
\kappa _{5}~ t]
\end{equation}
with
$$\xi =\frac{3(\gamma-\beta-1)^{2}}{\gamma^{2}}
+\frac{\kappa_{5}^{2}\Lambda_{5}}{2H^{2}}\left[1-\frac{(\gamma-\beta-1)^{2}}
{\gamma^{2}}\right]$$

The expressions for different physical parameters are

\begin{equation}\left.
\begin{array}{llll}
H =\frac{\theta}{3}= \sqrt{\frac{\Lambda_{5}}{6}} ~\kappa _{5}
~\text{coth} [(1+\beta) \sqrt{\frac{3\Lambda _{5}}{2}}~ \kappa
_{5}~ t]
\\\\
\rho = \frac{\sqrt{6 \Lambda _{5}}~(1+\beta)}{\kappa _{5}\gamma}~
\text{cosech} [ \sqrt{ \frac{3 \Lambda _{5}}{2}} ~\kappa _{5} (1+
\beta)~ t]
\\\\
q = 3 (1+ \beta) ~\text{sech}^{2}[ \sqrt{ \frac{3 \Lambda
_{5}}{2}}~ \kappa _{5} (1+ \beta)~ t] -1
\\\\
\lambda = \frac{(\gamma-\beta-1)\sqrt{6 \Lambda _{5}}}{\kappa
_{5}\gamma}~ \text{cosech} [ \sqrt{ \frac{3 \Lambda _{5}}{2}}~
\kappa _{5} (1+ \beta)~ t]
\\\\
\sigma=\sqrt{\frac{\sum K_{i}^{2}}{2}}~V^{-1}
\\\\
\Lambda =\frac{3(\gamma-\beta-1)^{2}}{\gamma^{2}}
H^{2}+\frac{\kappa_{5}^{2}\Lambda_{5}}{2}\left[1-\frac{(\gamma-\beta-1)^{2}}{\gamma^{2}}\right]
\\\\
G = \frac{\kappa_{5}^{3}(\gamma-\beta-1)}{\gamma}
\sqrt{\frac{\Lambda _{5}}{6}} ~\text{cosech} [ \sqrt{ \frac{3
\Lambda _{5}}{2}} ~\kappa _{5} (1+ \beta)~ t]
\end{array}\right\}
\end{equation}\\\

\begin{figure}
\includegraphics[height=1.7in]{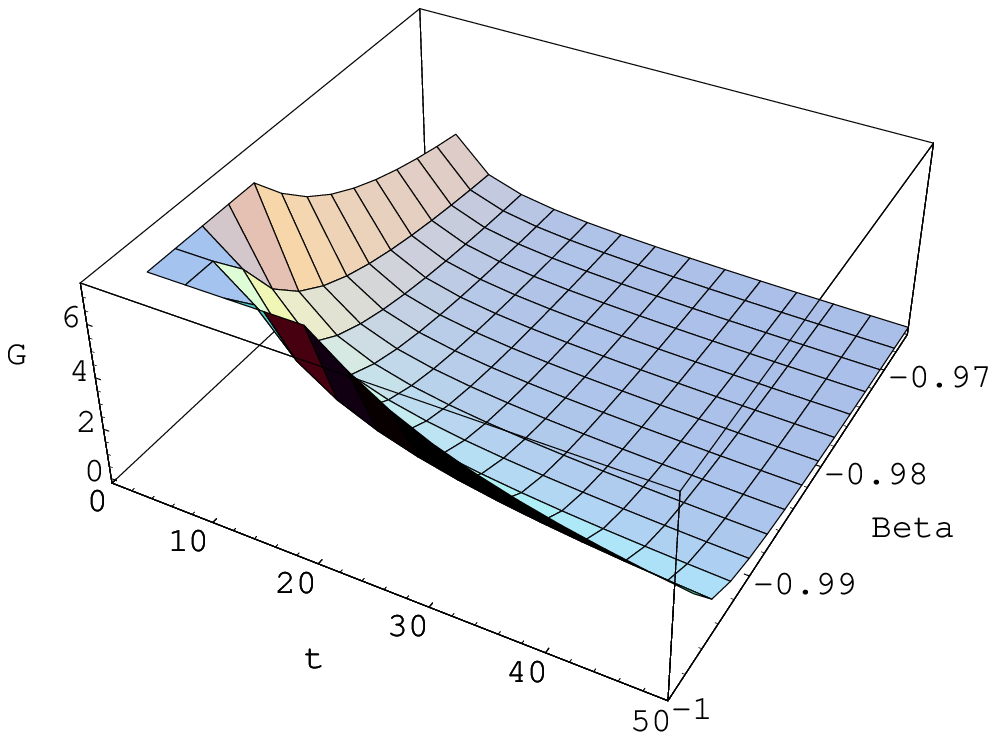}~~~
\includegraphics[height=1.7in]{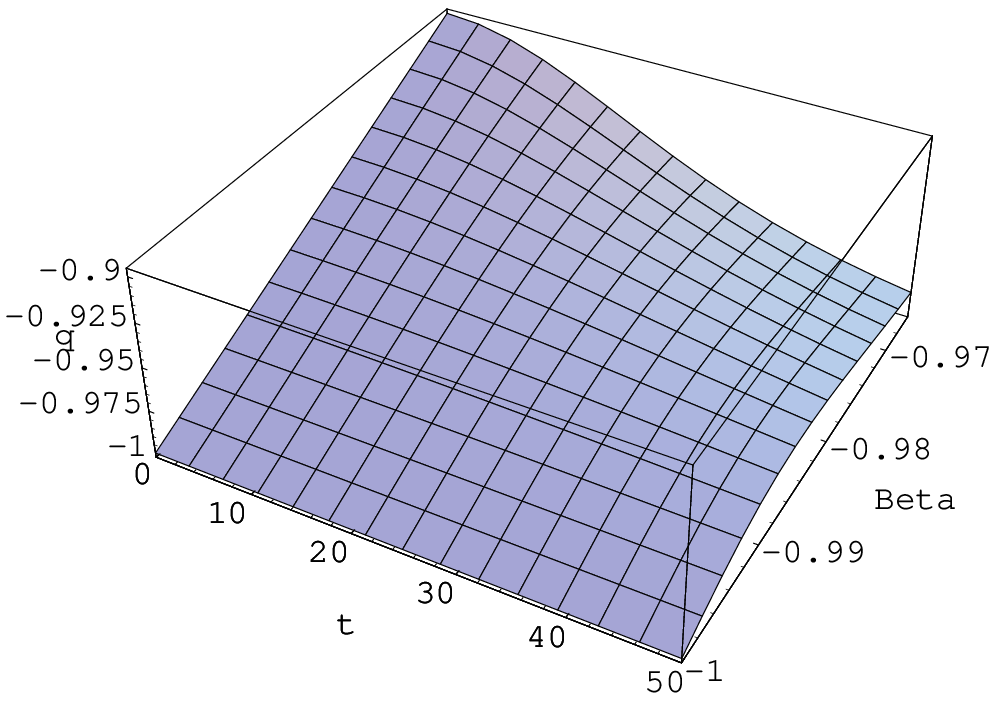}\\
\vspace{1mm}
Fig.1~~~~~~~~~~~~~~~~~~~~~~~~~~~~~~~~~~~~~~~~~~~Fig.2\\
\vspace{5mm} Figs. 1 shows variation of $G$ for $\Lambda_{5}>0$
with the choice $\sqrt{\frac{3\Lambda_{5}}{2}}~\kappa_{5}=1$ for
$\gamma= 4/3$. Fig. 2 presents the acceleration or deceleration
of the universe at different cosmic time for
$\Lambda_{5}>0$.\hspace{2cm} \vspace{6mm}

\includegraphics[height=1.7in]{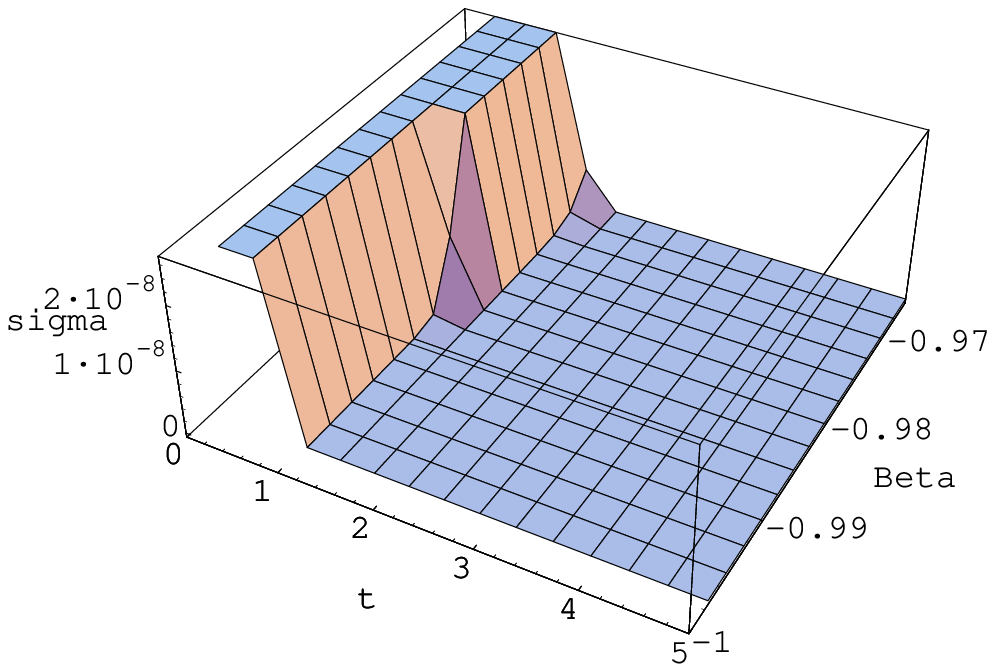}~~~
\includegraphics[height=1.7in]{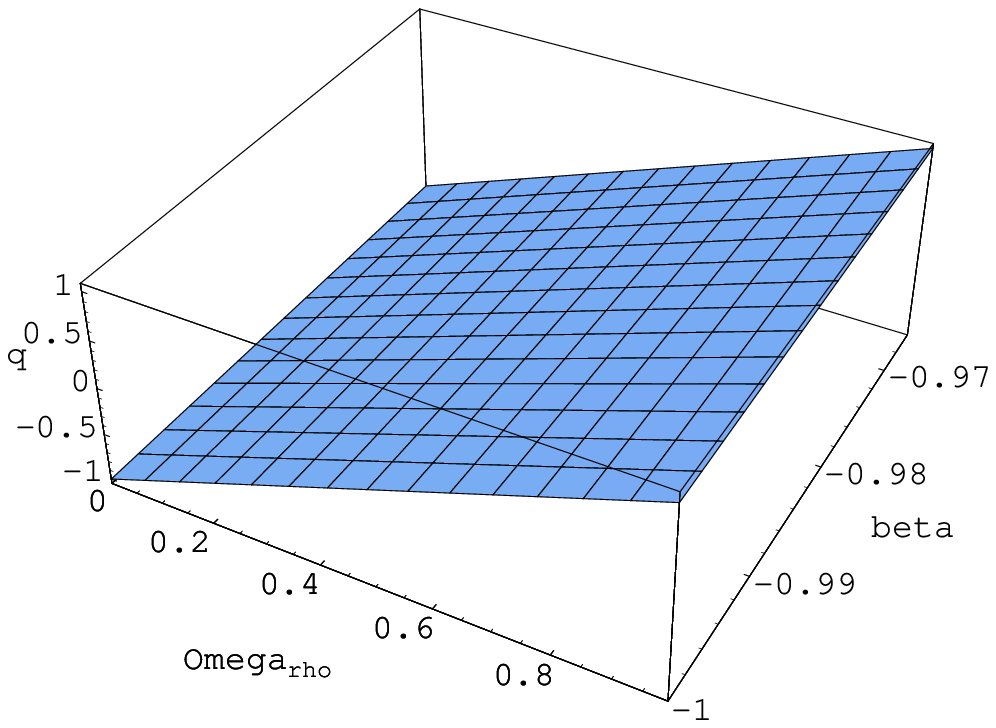}\\
\vspace{1mm}
Fig.3~~~~~~~~~~~~~~~~~~~~~~~~~~~~~~~~~~~~~~~~~~~Fig.4\\
\vspace{5mm}  Fig. 3 shows the gradual isotropization of the
universe with time. Fig. 4 gives the variation of $q$ with the
variation of $\Omega_{\rho}$ and $\beta$ for $\Lambda_{5}\ne 0$.
\hspace{2.9in} \vspace{6mm}

\end{figure}

From this solution we observe that  $H$ decreases  sharply  with
time for small $t$ and approaches a constant value  $ \kappa_{5}
\sqrt{\frac{\Lambda _{5}}{6}}$. This assures the positivity of
$\rho,~ \lambda ~\text{and}~ G $. Asymptotically, for large $t$,
the universe expands exponentially with positive acceleration $q
\approx -1$ and $\Lambda$ and $G$ become constant (see figs. 1 and
2) while universe isotropizes due to exponential fall off of the
anisotropic scalar (see fig. 3). Also the interdependence of
$\Omega_{\rho}$ and $~\beta$ over $q$ has been
shown in the 3D graph (fig. 4).\\

Further, the expressions for density parameter and the age of the
universe are

$$\Omega _{\rho} = \frac{2(\gamma-\beta-1)(1+q)}{3\gamma^{2}}$$
\begin{equation}
T=\frac{1}{(1+\beta)\sqrt{a}}~\text{tanh}^{-1}\left[\frac{2+3\beta-q}{3(1+\beta)}\right]^{1/2}
\end{equation}

So we can write
\begin{equation}
\beta =\gamma-1-\frac{3\Omega _{\rho}\gamma^{2}}{2(1+q)}
\end{equation}
i.e., $\beta$ can be estimated from the observed values.\\

Again from the relations (37) we have
\begin{equation}
G=\frac{\kappa^{2}_{5}(\gamma-\beta-1)\sqrt{1+q}}{\gamma\sqrt{3(1+\beta)}}H
\end{equation}
Also from (30) using (37) we write
\begin{equation}
a=\frac{3}{2}~\kappa^{2}_{5}\Lambda_{5}=\frac{3H^{2}(2+3\beta-q)}{1+\beta}
\end{equation}
Hence from the above two relations we can evaluate the five
dimensional quantities $\kappa_{5}$ and $\Lambda_{5}$. Thus the
solution contains no free parameter.\\

We note that a non-vanishing cosmological constant in the bulk
induces a natural time scale in $4D$ as $$\tau _{s} = \sqrt{
\frac{6}{\kappa ^{2}_{5} \Lambda_{5}}}$$ which in terms of
observed quantities has the expression
\begin{equation}
\tau
_{s}=\frac{1}{H}~\left(1-\frac{\gamma^{2}\Omega_{\rho}}{2(\gamma-\beta-1)(1+\beta)}\right)^{-1/2}
\end{equation}

If we consider a model universe with $\Omega_{\rho}\approx 0.3$
and that of $H\approx 0.7\times 10^{-10}~ y_{r}^{-1}$, then the
characteristic scale is $\approx 5.2\times 10^{10}~y_{r}s$ i.e.,
52 billion years, which is much greater than the age of the
universe (due to data from WMAP mission) [21].\\\

 Lastly, from the observational bound for $\beta$ the deceleration
parameter is restricted by $\frac{3\gamma\Omega _{\rho}}{2}-1<q <
\frac{3\Omega_{\rho}\gamma^{2}}{2(\gamma-.034)}-1$, which shows
that for $\Omega_{\rho}<\frac{2(\gamma-.034)}{3\gamma^{2}}$, there
is always an accelerating universe while for
$\Omega_{\rho}>\frac{2(\gamma-.034)}{3\gamma^{2}}$ both
acceleration and deceleration is possible (see fig. 4), but for
$\Omega_{\rho}>\frac{2}{3\gamma}$, only we have decelerating universe.\\\

Thus in the present matter dominated era ($\gamma=1$) to get
accelerated universe $\Omega_{\rho}<0.644$ which supports the
observed value of $\Omega_{\rho}$. Therefore it is possible to
have at present an accelerating universe in this model.\\\\

{\bf Case II}:  ~~$\Lambda _{5} ~(\text{i.e.,}~ a) =0$\\

Here the relation (34) should be satisfied and $V$ has the simple
solution
\begin{equation}
V=\left[\sqrt{b}(\beta+1)t\right]^{\frac{1}{\beta+1}} ,~~(\beta+1
\neq 0)
\end{equation}
with $$\xi=\frac{3(\gamma-1-\beta)^{2}}{\gamma^{2}}$$ (Note that
$\beta+1 \neq 0$; otherwise there is no variation of $G$ )\\

Here also the integration constant $t_{0}$ is taken to be zero
for simplicity. The scale factors will have an explicit forms as
$$a_{i}=a_{i0} V^{\frac{1}{3}}~ \text{exp}
\left[-\frac{K_{i}}{3\beta\sqrt{b}}~V^{\beta}\right] ,~~i=1,2,3$$

The deceleration  parameter will have constant value and is
related to $\beta$ by the relation $$q=2+3\beta$$

The brane tension, $4D$ effective cosmological term  $\Lambda$
and the gravitational coupling $G$ has the time variation
as\\\
\begin{equation}
\lambda=\frac{2(\gamma-1-\beta)}{\kappa^{2}_{5}\gamma(\beta+1)t},~~~
\Lambda=\frac{3(\gamma-1-\beta)^{2}}{\gamma^{2}}H^{2},~~~
G=\frac{\kappa^{2}_{5}(\gamma-1-\beta)H}{\gamma}
\end{equation}
with
$$H=\frac{1}{3(1+\beta)t}$$ as Hubble parameter.\\

\begin{figure}

\includegraphics[height=2in]{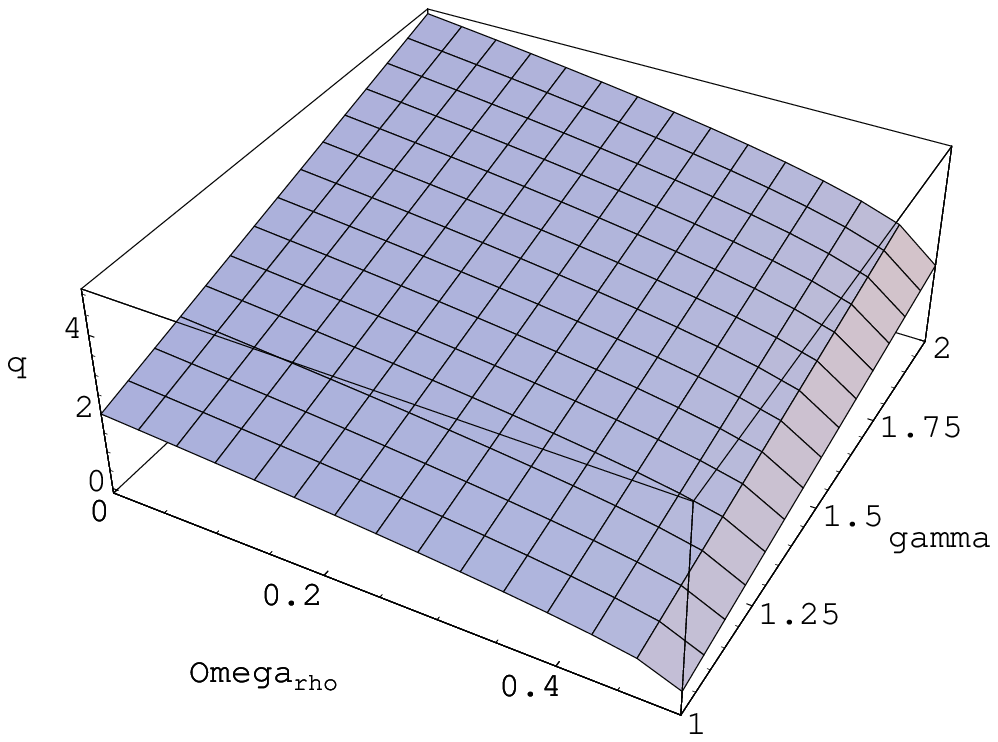}~~~
\includegraphics[height=2in]{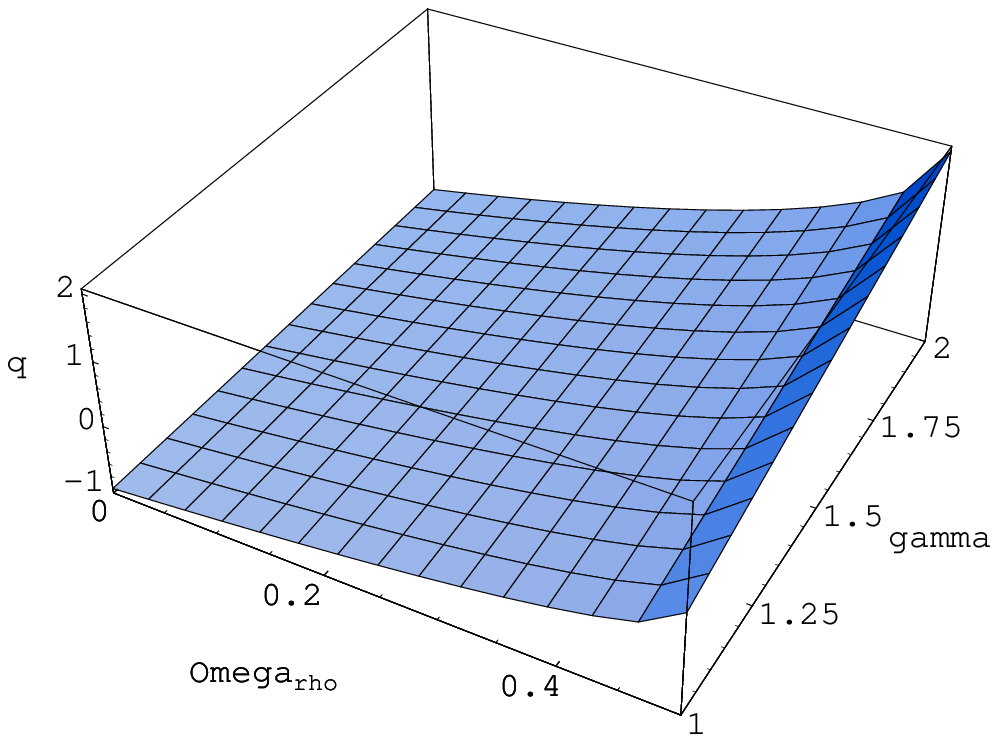}\\
\vspace{1mm}
Fig.5~~~~~~~~~~~~~~~~~~~~~~~~~~~~~~~~~~~~~~~~~~~Fig.6\\
\vspace{5mm} Figs. 5 and 6 present the variation of $q$ due to the
variation of $\Omega_{\rho}$ and $\gamma$ for `$+$' and `$-$'
sign in equation (46).\hspace{4.5in} \vspace{6mm}

\end{figure}

The anisotropy scalar $\sigma$ decreases with time as
$t^{-\frac{1}{1+\beta}}$ and vanishes at infinity. Now for energy
density namely $\rho =\frac{2}{\kappa^{2}_{5}\gamma t}$~, we have
the following relation among density parameter $\Omega_{\rho}$,
$\beta$ and $\gamma$

\begin{equation}
\frac{
G\rho}{3H^{2}}=\Omega_{\rho}=\frac{2(\gamma-\beta-1)(\beta+1)}{\gamma^{2}}
\end{equation}

which gives $\beta$ as a function of $\gamma$ and $\Omega_{\rho}$

\begin{equation}
\beta+1=\frac{\gamma}{2}~[1\pm \sqrt{1-2\Omega_{\rho}}]
\end{equation}

Thus for real $\beta$ we must have $\Omega_{\rho}\le 1/2$. This
upper limit of $\Omega_{\rho}$ is independent of any specific
values of $\gamma$ and $q$. Also eliminating `$\beta$' between
(45) and the expression for $q$ we have

\begin{equation}
\Omega_{\rho}=\frac{2(3\gamma-q-1)(1+q)}{9\gamma^{2}}
\end{equation}

The variation of $q$ over $\Omega_{\rho}$ and $\gamma$ has been
shown in figures 5 and 6 for positive and negative signs in
equation (46). In the first case there is always deceleration
while in the second case $q$ varies between $-1$ and 2.\\

Now the relationship between the density parameters are

\begin{equation}
\frac{\Lambda}{3H^{2}}=\Omega_{\Lambda}=\frac{9\gamma^{2}\Omega_{\rho}^{2}}{4(1+q)^{2}}
\end{equation}

Further we note that the above solution can be recovered from the
solution for $\Lambda_{5}>0$ if we proceed to the limit
$\Lambda_{5}\rightarrow 0$. Near the big bang epoch (i.e., for
small $t$) the present solution has similar behaviour with that
for positive $\Lambda_{5}$. Lastly, to have an accelerated
universe the density parameter should have further restriction on
upper bound (depending on $\gamma$), considering `$-$' sign in
equation (46) as
\begin{equation}
\Omega_{\rho}<\frac{2(3\gamma-1)}{9\gamma^{2}}
\end{equation}

We note that this upper bound for $\Omega_{\rho}$ is always less
that 0.5, hence it is possible to have a decelerating phase of
the universe when $\Omega_{\rho}$ lies between the bound in
equation (49) and 0.5. On the other hand, for `$+$' sign in
equation (46) the universe will accelerate if $\gamma<2/3$ and
then there is a lower bound for $\Omega_{\rho}$ i.e.,

\begin{equation}
\frac{2(3\gamma-1)}{9\gamma^{2}}<\Omega_{\rho}\le 0.5
\end{equation}

To have an end of this section we remark that similar to the
previous solution for $\Lambda_{5}>0$, here also there is no free
parameter remain, all of them can be estimated from observation
(i.e., $\Omega_{\rho}$ and $H$). In table I and II we have
presented the numerical values of different parameters based on
various values of the density parameter $\Omega_{\rho}$ for dust
model ($\gamma=1$) with $H=0.7\times 10^{-10}$ yr$^{-1}$. Table I
corresponds to `$-$' sign in the expression for $\beta$ in
equation (46) while table II for `$+$' sign. In table I,
throughout there is acceleration except for $\Omega_{\rho}=0.5$
as expected from the expression for $q$. In fact, for
$\Omega_{\rho}>0.44$ only deceleration is possible. We note that
here $\Omega_{\rho}+\Omega_{\Lambda}\ne 1$, there is a
contribution of $\Omega_{\rho^{2}}$ in brane scenario. But this
contribution is insignificant in the present epoch. Thus the data
in this table are consistent with recent observation as there is
deceleration followed by acceleration. On the otherhand, table II
has the reverse picture. There is always deceleration as
$\gamma=1$ and $\Omega_{\rho^{2}}$ has significant contribution
in the matter content of the universe. Finally, we say that table
II represents early phases of the universe when brane scenario
has dominant contribution while table I is valid in the present
epoch when brane effect is negligible.

$$\text {TABLE-I}$$ ~~~~~~~~~~~~~~~~~\text{($\Lambda_{5}=0$. Present phase of the
evolution, accelerated expansion)}
\[
\begin{tabular}{|l|l|l|r|r|r|}
\hline\hline \multicolumn{1}{|c|}{~~~$\Omega_{\rho}$~~~} &
\multicolumn{1}{c|}{~~~~$\beta$~~~~} &
\multicolumn{1}{c|}{~~~~$q$~~~~} & \multicolumn{1}{c|}{~~~~$\Omega_{\Lambda}$~~~~} & \multicolumn{1}{c|}{~~$T/10^{10}$~yr's~~} \\
\hline
&  &  &  &  \\
~0.5 &  $-$0.5 &  ~~~0.5 & 0.25~~~  & 0.95238 ~~~~~\\
\hline
&  &  &  &  \\
~0.4 &  $-$0.72361 &  $-$0.17082 & 0.52361  & 1.72287~~~~~~\\
\hline
&  &  &  &  \\
~0.3 &  $-$0.81623 &  $-$0.44868 & 0.66623  & 2.59120 ~~~~~\\
\hline
&  &  &  &  \\
~0.2 &  $-$0.88730 &  $-$0.66189 & 0.7873  & 4.22523 ~~~~~\\
\hline
&  &  &  &  \\
~0.1 &  $-$0.94721 &  $-$0.84164 & 0.89721  & 9.02108~~~~~~\\
\hline
&  &  &  &  \\
~0.08 &  $-$0.95826 &  $-$0.87477 & 0.91826  & 11.4078 ~~~~~\\
\hline
&  &  &  &  \\
~0.04 &  $-$0.97958 &  $-$0.93875 & 0.95958  & 23.3234 ~~~~~\\

\hline\hline
\end{tabular}%
\]\\

\newpage
$$\text {TABLE-II}$$ ~~~~~~~~~~~~~~~~~\text{($\Lambda_{5}=0$. Early phase of the evolution,
decelerated expansion)}
\[
\begin{tabular}{|l|l|l|r|r|r|}
\hline\hline \multicolumn{1}{|c|}{~~~$\Omega_{\rho}$~~~} &
\multicolumn{1}{c|}{~~~~$\beta$~~~~} &
\multicolumn{1}{c|}{~~~~$q$~~~~} & \multicolumn{1}{c|}{~~~~$\Omega_{\Lambda}$~~~~} & \multicolumn{1}{c|}{~~$T/10^{10}$~yr's~~} \\
\hline
&  &  &  &  \\
~0.5 &  $-$0.5 &  ~~~0.5 & 0.25~~~  & 0.95238 ~~~~~\\
\hline
&  &  &  &  \\
~0.4 &  $-$0.27639 &  ~~1.17082 & 0.07639  & 0.65808~~~~~~\\
\hline
&  &  &  &  \\
~0.3 &  $-$0.18377 &  ~~1.44868 & 0.03377  & 0.58340 ~~~~~\\
\hline
&  &  &  &  \\
~0.2 &  $-$0.11270 &  ~~1.66190 & 0.01270  & 0.53668 ~~~~~\\
\hline
&  &  &  &  \\
~0.1 &  $-$0.05279 &  ~~1.84164 & 0.00279  & 0.50273~~~~~~\\
\hline
&  &  &  &  \\
~0.08 &  $-$0.04174 &  ~~1.87477 & 0.00174  & 0.49693 ~~~~~\\
\hline
&  &  &  &  \\
~0.04 &  $-$0.02042 &  ~~1.93875 & 0.00042  & 0.48612 ~~~~~\\

\hline\hline
\end{tabular}%
\]\\\\

{\bf Case III}:    ~~$\Lambda_{5} ~(\text{i.e.,}~ a)< 0$\\

This case is important from the point of view of brane scenario
as proposed by Randall and Sundrum [8] because according to them
a single brane is embedded in a five dimensional anti-de-Sitter
bulk (AdS$_{5}$).

For this choice the solution of $V$ can be written as (with
$t_{0}=0$)
\begin{equation}
V ^{1+\beta}=\sqrt{\frac{b}{|a|}} ~\text{sin}[(1+\beta)\sqrt{|a|}~
t]
\end{equation}
with the expressions for the physical parameters as
\begin{equation}\left.
\begin{array}{llll}
H =\frac{\theta}{3}=\sqrt{|a|}~\text{cot} [(1+ \beta) \sqrt{|a|}~
t]
\\\\
\rho =\frac{\sqrt{6  |\Lambda_{5}|}~(1+\beta)}{\kappa_{5}\gamma}
~\text{cosec} [(1+ \beta) \sqrt{|a|}~ t]
\\\\
q =  3(1+ \beta) ~\text{sec} ^{2}[(1+ \beta) \sqrt{|a|}~ t] -1
\\\\
\lambda = \frac{(\gamma-\beta-1)\sqrt{6
|\Lambda_{5}|}}{\kappa_{5}\gamma}~\text{cosec}[(1+ \beta)
\sqrt{|a|}~ t]
\\\\
\sigma=\sqrt{\frac{\sum K_{i}^{2}}{2}}~V^{-1}
\\\\
\Lambda =\frac{3(\gamma-\beta-1)^{2}}{\gamma^{2}}
H^{2}+\frac{\kappa_{5}^{2}\Lambda_{5}}{2}\left[1-\frac{(\gamma-\beta-1)^{2}}{\gamma^{2}}\right]
\\\\
G = \frac{\kappa_{5}^{3}(\gamma-\beta-1)}{\gamma}
\sqrt{\frac{|\Lambda _{5}|}{6}}~\text{cosec}[(1+ \beta)
\sqrt{|a|}~ t]
\end{array}\right\}
\end{equation}

This solution represents a spatially flat but recollapsing
universe with recollapsing  time ($\tau _{rec}$) given by
$$ \text{sin} [(1+\beta) \sqrt{|a|}~\tau _{rec}]=0$$

Though this solution is completely different from the solution
with positive $\Lambda _{5}$ but we can obtain this solution from
(37) if we change $ \sqrt{\Lambda _{5}} \rightarrow  i
\sqrt{|\Lambda _{5}|}$. As a consequence, $\sqrt{a} \rightarrow
i\sqrt{|a|}$ and $\text{sinh}(\sqrt{a}~t) \rightarrow
i~\text{sin}(\sqrt{|a|}~t)$. Finally, the age of the universe can
be written as
$$
\tau=\frac{\tau_{rec}}{\pi}~\text{tan}^{-1}\left(\frac{q-2-3\beta}{3(1+\beta)}\right)^{1/2}
$$

\section{\normalsize\bf Bianchi V model}
The Bianchi V space-time is an isotropic generalization of the
open ($k=-1$) Robertson-Walker geometry and the metric-ansatz has
the expression

\begin{equation}
ds ^{2}=-dt ^{2}+a _{1}^{2}(t)dx ^{2} +a_{2}^{2}(t) e^{-2x} dy
^{2} +a _{3}^{2}(t) e^{-2x}dz^{2}
\end{equation}

As before the field equations are equation (20) and
\begin{equation}
\frac{1}{V}\frac{d}{dt}(VH_{i}) -2a_{1}^{-2}=\Lambda-
\frac{(\gamma-2)}{2}\kappa ^{2}\rho -\frac{(\gamma-1)}{12}\kappa
^{4}_{5}\rho ^{2}+\frac{U_{0}}{V ^{\frac{4}{3}}},~i=1,2,3
\end{equation}
\begin{equation}
2H_{1} -H_{2}-H_{3}=0
\end{equation}

Now compare to the Bianchi I model in the previous section, we
have the same relation (28) among the Hubble parameters but the
constants are given by
\begin{equation}
K_{1}=0,~ K_{2}= - K_{3}
\end{equation}

due to the field equation (46). The differential equation for $V$
in the present model is of the form (after integrating once)
\begin{equation}
\dot{V}^{2}=a V^{2}+b V^{-2\beta}+c V^{\frac{2}{3}}
-\frac{4}{V}+d_{1}
\end{equation}

Also as before considering the variation of $\Lambda$ as
$\Lambda=\xi H^{2}$, we have the same differential equation (34)
as in Bianchi I model. Comparing (48) with (34) we note that
Bianchi V model is possible only for $\beta=\frac{1}{2}$ (with
$c=0=d_{1}$) which contradicts the observational bounds on
$\beta$. The explicit solution is (choosing the integration
constants appropriately)

\begin{equation}t=\left\{
\begin{array}{llll}
\frac{2}{3\sqrt{a}}~\text{cosh}
^{-1}V^{\frac{3}{2}}\sqrt{|\frac{a}{b-4}|}, ~~\text{if}~~
\frac{b-4}{a}<0\\\\
\frac{2}{3\sqrt{a}}~\text{sinh}
^{-1}V^{\frac{3}{2}}\sqrt{|\frac{a}{b-4}|}, ~~\text{if}~~
\frac{b-4}{a}>0\\\\
\frac{2}{3\sqrt{a}}~\text{log} V^{3/2}, ~~\text{if}~~ b=4\\\\
\end{array}
\right.
\end{equation}

The explicit form for the physical parameters are : \\

\begin{equation}\left.
\begin{array}{llll}
H = \frac{\sqrt{a}}{3}~\text{tanh} (\frac{3}{2} \sqrt{a}~ t)
\\\\
\rho =D \sqrt{\frac{a}{b-4}}~(\text{cosh} \frac{3}{2} \sqrt{a}~
t)^{-1}
\\\\
q = - 1 -\frac{9}{2}~\text{cosech} ^{2}(\frac{3}{2} \sqrt{a}~ t)
\\\\
\lambda = \frac{2}{3} D (\gamma
-\frac{3}{2})\sqrt{\frac{a}{b-4}}~(\text{cosh} \frac{3}{2}
\sqrt{a}~ t)^{-1}
\\\\
\Lambda =\xi H ^{2}
\\\\
G \propto (\text{cosh} \frac{3}{2} \sqrt{a}~ t)^{-1}
\end{array}\right\}~~~ \text{for}~~ \frac{b-4}{a} <0
\end{equation}

\begin{equation}\left.
\begin{array}{llll}
H =\frac{\sqrt{a}}{3}~\text{coth}(\frac{3}{2} \sqrt{a}~ t)
\\\\
\rho =D \sqrt{\frac{a}{b-4}}~(\text{sinh} \frac{3}{2} \sqrt{a}~
t)^{-1}
\\\\
q = - 1 +\frac{9}{2}~\text{sech} ^{2}(\frac{3}{2} \sqrt{a}~ t)
\\\\
\lambda = \frac{2}{3} D (\gamma
-\frac{3}{2})\sqrt{\frac{a}{b-4}}~(\text{sinh} \frac{3}{2}
\sqrt{a}~ t)^{-1}
\\\\
\Lambda =\xi H ^{2}
\\\\
G \propto (\text{sinh} \frac{3}{2} \sqrt{a}~ t)^{-1}
\end{array}\right\}~~~ \text{for}~~ \frac{b-4}{a} >0
\end{equation}
\\

\begin{equation}\left.
\begin{array}{llll}
H = \frac{\sqrt{a}}{3}
\\\\
\rho =D ~e^{-\frac{3}{2}\sqrt{a}~t}
\\\\
q =  -1
\\\\
\lambda = \frac{2}{3}D (\gamma
-\frac{3}{2})~e^{-\frac{3}{2}\sqrt{a}~t}
\\\\
\Lambda =\xi H^{2} \\\\
G \propto e^{-\frac{3}{2}\sqrt{a}~t}
\end{array}\right\}~~~\text{for}~~b=4
\end{equation}\\

\section{\normalsize\bf Discussion}

In this paper, we have studied the consequences of the variation
of $G$ and $\Lambda$ in the form $\frac{\dot{G}}{G}\sim H$ and
$\Lambda \sim H ^{2}$ on Bianchi I and V cosmological models
based on the brane-world scenario. We have obtained solutions in
Bianchi I model for positive, zero and negative value of the
cosmological constant $\Lambda _{5}$ on the bulk. From the
solution with positive $\Lambda _{5}$ one can obtain solution for
negative $\Lambda _{5}$ by analytic continuation $(\sqrt{\Lambda
_{5}} \rightarrow i \sqrt{|\Lambda _{5}|})$ while solution for
$\Lambda _{5} =0$ are the limiting solution of (37) as
$\Lambda_{5}\rightarrow 0$. However, asymptotically, as $t
\rightarrow \infty$ the nature of the solutions in three cases
are distinct. For $\Lambda _{5} >0$, the universe expands
exponentially with constant $\Lambda $ as $t \rightarrow \infty$,
while for $\Lambda _{5} =0$ the universe expands in a power law
fashion (for $\beta \neq -1$) with a constant deceleration and
all physical parameters become vanishingly small. But for
$\Lambda _{5} < 0$, the solution is an expanding-collapsing model
of the universe i.e., an oscillatory model with finite time
period.\\

\begin{figure}

\includegraphics[height=2in]{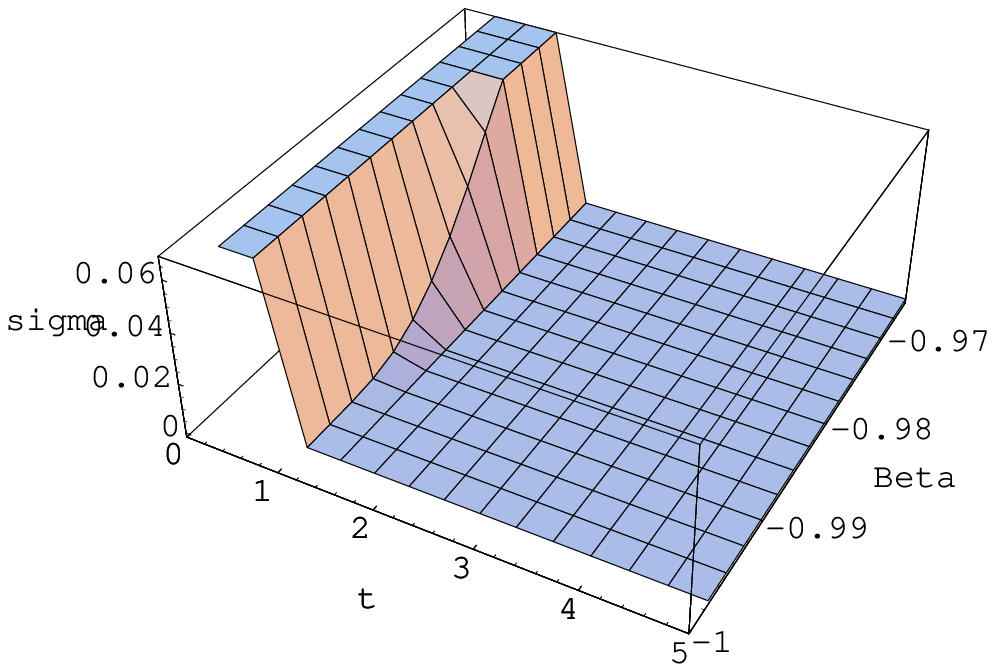}\\
\vspace{1mm}
Fig.7\\
\vspace{5mm} Fig. 7 showing $\sigma$ against $t$ and $\beta$
where $\sigma=t^{-\frac{\beta+2}{\beta+1}}$ for standard cosmology
Bianchi I model.\hspace{13cm}\vspace{6mm}

\end{figure}

For $\Lambda_{5}>0$, $H$ and $G$ fall off sharply from large
positive value for small $t$ to a constant value for large $t$,
while $\Lambda$ approaches to zero asymptotically. Also for
$\Lambda_{5}=0$, the behaviour of these parameters are similar
but all of them vanish asymptotically. But we have peculiar
behaviour of these parameters for negative $\Lambda_{5}$ as there
are points of discontinuities at finite $t$. We have also
presented solutions for Bianchi V model with $\Lambda_{5}>0$ only
for different signs (including zero) of the constant
$\frac{b-4}{a}$, but they are not realistic as they corresponds
to $\beta=0.5$, which is outside the observational bounds on
$\beta$. Their mathematical form is very similar to the Bianchi I
case, with one correction term, proportional to $V^{4/3}$ added
to the parametric time equation. Hence the general physical
behaviour in both geometries has the similar qualitative
features. Lastly, we have shown the behaviour of the shear scalar
($\sigma\propto t^{-\frac{\beta+2}{\beta+1}}$) against $t$ and
$\beta$ in fig.7 for standard cosmology Bianchi I model. In
standard model $\sigma$ falls off as polynomial in `$t$' while in
brane scenario it falls exponentially. In fact the figures 3 and
7 show the similarity of the anisotropy scalar in brane scenario
and in standard
cosmological model.\\

Finally, we like to emphasize that the whole analysis of this
paper is independent of any particular choice of the bulk
parameters which is a good characteristic for the brane-world
models.\\\\

{\bf Acknowledgement:}\\\\
The authors are grateful to the referee for his valuable
suggestions. U.D. is thankful to C.S.I.R., Govt. of  India  for
awarding  a Senior  Research Fellowship.\\

{\bf References:}\\
\\
$[1]$  V. N. Melnikov, ``{\it Gravity as a key Problem of the Millennium}'',
Proc. 2000 NASA/JPL Conference on Fundamental Physics in Microgravity, CD-Version,
NASA Document D-21522, 4.1-4.7 (2001) (Solvang, CA, USA), {\it gr-qc}/0007067.\\
$[2]$ P. A. M. Dirac, {\it Nature} {\bf 139} 323 (1937); {\it
Proc. Royal Soc. London} {\bf 165} 199 (1938).\\
$[3]$ D. W. Sciama, {\it Mon. Not. R. Astron. Soc.} {\bf 113} 34
(1953).\\
$[4]$ P. Jordan, Scherkraft and Weltall, Friedrich
Vieweg and Sohn, Braunschwelg (1955).\\
$[5]$ C. Brans and R. Dicke, {\it Phys. Rev.} {\bf 124} 925
(1961).\\
$[6]$ S. Perlmutter et al, {\it Astrophys. J.} {\bf 483} 565
(1997).\\
$[7]$ Y. Fujii, {\it Gravitation and Cosmology} {\bf 6} 107
(2000).\\
$[8]$ M. Gasperini, {\it Phys. Lett. B} {\bf 194} 347 (1987); {\it Class. Quantum Grav.} {\bf 5} 521 (1988);
K. Freese, F.C. Adams, J.A. Friemann and E. Mottolla, {\it Nucl. Phys. B} {\bf 287} 797 (1987); P.J.E. Peebles and
B. Ratra, {\it Astrophys. J} {\bf 325} L17 (1988); W. Chen and Y.S. Wu, {\it Phys. Rev. D} {\bf 41} 695 (1990).\\
$[9]$ S. Perlmutter et al, {\it Nature} (London) {\bf 391} 51 (1998);
R.R. Caldwell, R. Dave and P.J. Steinhardt, {\it Phys. Rev. Lett.} {\bf 80}
1582 (1998); L.P. Chimento, A. S. Jakubi and D. Pavon, {\it Phys. Rev. D} {\bf 62} 063508 (2000).\\
$[10]$ A. Bonanno and M. Reuter, {\it Phys. Lett.} {\bf 527B} 8 (2002).\\
$[11]$I.L. Shapiro, J. Sola, C. Espana-Bonnet and P. Ruiz-Lapnente, {\it Phys. Lett.} {\bf 574B} 149 (2003).\\
$[12]$ R. G. Vishwakarma, {\it Class. Quantum Grav.} {\bf 19} 4747
(2002).\\
$[13]$ L. Randall and R. Sundrum, {\it Phys. Rev. Lett.} {\bf 83}
3370 (1999); {\it Phys. Rev. Lett.} {\bf 83} 4690 (1999).\\
$[14]$ K. Maeda and D. Wands, {\it Phys. Rev. D} {\bf 62} 124009
(2000).\\
$[15]$ M. Sasaki, T. Shiromizu and K. Maeda, {\it Phys. Rev. D}
{\bf 62} 024008 (2000); T. Shiromizu, K. Maeda and M. Sasaki,
{\it Phys. Rev. D} {\bf 62} 024012 (2000)\\
$[16]$ R. Maartens, {\it Phys. Rev. D} {\bf 62} 084023 (2000).\\
$[17]$ R. Maartens, ``{\it Geometry and Dynamics of
Brane-World}'', {\it gr-qc}/0101059.\\
$[18]$ C. M. Chen, T. Harko and M. K. Mak, {\it Phys. Rev. D}
{\bf 64} 044013 (2001); T. Harko and M. K. Mak, {\it Class. Quantum Grav.} {\bf 20} 407 (2003).\\
$[19]$ J. Ponce de Leon, ``{\it Brane Universes with Variable $G$
and $\Lambda_{(4)}$}'', {\it Class. Quantum Grav.} {\bf 20} 5321 (2003); ({\it gr-qc}/0305041).\\
$[20]$ J. Ponce de Leon, {\it Mod. Phys. Lett. A} {\bf 17} 2425
(2002); {\it Int. J. Mod. Phys. D} {\bf 11} 1355 (2002).\\
$[21]$ D. N. Spergel et al, ``{\it First Year Wilkinson Microwave
Anisotropy Probe (WMAP) Observations: Determination of
Cosmological Parameters}'', {\it astro-ph}/0302209.\\

\end{document}